\documentclass[pdflatex,sn-mathphys-num]{sn-jnl}
\usepackage[switch]{lineno}
\usepackage{graphicx}%
\usepackage{multirow}%
\usepackage{amsmath,amssymb,amsfonts}%
\usepackage{amsthm}%
\usepackage{mathrsfs}%
\usepackage[title]{appendix}%

\usepackage{textcomp}%
\usepackage{manyfoot}%
\usepackage{booktabs}%
\usepackage[utf8]{inputenc}
\usepackage{newunicodechar}
\newunicodechar{∼}{\textasciitilde{}}
\usepackage{algorithm}%
\usepackage{algorithmicx}%
\usepackage{algpseudocode}%
\usepackage{listings}%
\usepackage{siunitx}
\let\orcidlogo\relax
\usepackage{orcidlink}

\begin{document}
%\linenumbers

\makeatletter
\renewcommand\@cite[2]{\textsuperscript{#1\ifx\\#2\\\else, #2\fi}}
\makeatother

\title{Beta-like tracks in a cloud chamber from nickel cathodes after electrolysis}

\author*[1]{\fnm{Shyam Sunder} \sur{Lakesar}\orcidlink{0000-0001-8685-5621}}\email{sslake37@gmail.com}

\author[2]{\fnm{Raj Ganesh} \sur{S. Pala}\orcidlink{0000-0001-5243-487X}}\email{rpala@iitk.ac.in}

\author[1]{\fnm{K P} \sur{Rajeev}\orcidlink{0000-0002-4685-6766}}\email{kpraj@iitk.ac.in}

\affil*[1]{\orgdiv{Department of Physics}, \orgname{Indian Institute of Technology}, \orgaddress{\city{Kanpur}, \country{India}}}

\affil[2]{\orgdiv{Department of Chemical Engineering}, \orgname{Indian Institute of Technology}, \orgaddress{\city{Kanpur}, \country{India}}}

\abstract{Electrochemically induced nuclear activity in hydrogen and deuterium-absorbing metals has been reported intermittently, yet a direct observation of nuclear signatures remains challenging. We electrolyzed light water with nickel cathodes under half-wave rectified RMS potentials of 5 V and 20 V and subsequently analyzed them using a Peltier-cooled diffusion-type Wilson cloud chamber for particle emission. The reacted cathodes emitted $\beta$-like particles forming condensation tracks of lengths of 0.6–16 mm and an average activity {\mathversion{normal}$0.6 \pm 0.1$} counts per minute (cpm) for 5~V samples and {\mathversion{normal}$1.0 \pm 0.1$} cpm for 20~V samples. No such emissions were detected from unreacted samples. These results provide empirical evidence that electrochemical reactions can generate radioactive isotopes in condensed matter.}


\maketitle

\section{Introduction}

Nuclear fission was first identified in condensed matter containing naturally radioactive isotopes.
More recently, reports of nuclear signatures in electrochemically loaded metal-hydride and metal-deuteride systems
have renewed interest in how condensed-matter environments may influence nuclear processes.
Experimental observations include energetic particle emissions
\cite{pines2020nuclear,czerski2024indications,mosier2017detection,roussetski2017detection}
as well as isotopic shifts \cite{huang2024water}.
Together, these results suggest that certain electrochemical conditions may enhance nuclear reaction probabilities
at energies far below those expected from conventional nuclear theory.
In standard models, hydrogen nuclei require kinetic energies in the hundreds of keV to MeV range
to overcome the Coulomb barrier.
In contrast, experimental studies have reported nuclear signatures in metal hydrides
under electrochemical conditions that correspond to characteristic energies of the order of a few eV \cite{huke2008enhancement}.

Strong cathodic potentials drive a substantial amount of deuterium (D) or hydrogen (H) into metallic lattices and increase the local reactant density\cite{green1994electrolytic}. Additional excitations can arise from plasma bombardment\cite{chen2025electrochemical,mckeown2025low} or acoustic cavitation\cite{huang2024water}. Despite extensive investigations, the direct detection of nuclear products in these experiments remains challenging due to the inherently low emission rates and lack of direct evidence. To address these limitations, we employ a custom-built Peltier-cooled diffusion Wilson cloud chamber (cloud chamber) for detecting charged-particle emissions from electrochemical nickel-hydrogen (Ni/H) systems. This approach enables direct observation of particle trajectories and allows the extraction of particle energies.

Although Pd/D systems have historically dominated studies of nuclear effects in metal lattices, Ni/H systems have shown comparable evidence of neutron emission\cite{battaglia1999neutron} and tritium($^{3}$H) formation\cite{notoya1994tritium}. Here, we report beta-like particle emissions from light water electrolyzed Ni cathodes under low voltage conditions. These results demonstrate that electrolysis can lead to the formation of trace amounts of radioactive isotopes, which are detectable and measurable using a cloud chamber. Given the direct evidence, it might be worthwhile if more comprehensive efforts are directed toward exploring the potential of electrochemically activated nuclear reactions in the areas of energy generation,  production of valuable radioactive isotopes, and radioactive waste treatment.

\section{Methods}
\begin{figure}[H]
    \centering
    \includegraphics[width=1\linewidth]{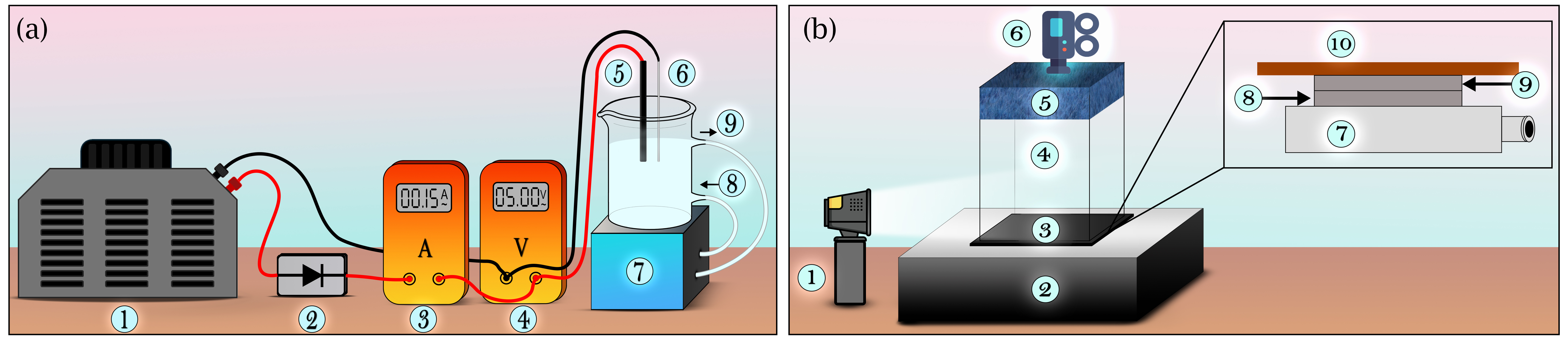}
    \caption{
(a) Schematic of the electrolytic cell setup: (1) autotransformer, (2) half-wave rectifier (HWR), (3) ammeter, (4) voltmeter, (5) graphite anode (99.99\% carbon, 6 mm diameter), (6) nickel cathode (99.5\% nominal Ni, 1 mm diameter), (7) automated electrolyte-refilling system containing electrolyte reservoir and pump, (8) electrolyte inlet, and (9) overflow outlet.\\
(b) Schematic of the cloud chamber: (1) projector light for track illumination, (2) cooling reservoir housing, the Peltier cooling modules and power supplies, (3) Copper plate maintained at approximately $-40^{\circ}\mathrm{C}$ with a thin black vinyl sheet on top, (4) cuboid glass chamber, (5) isopropyl-alcohol-soaked felt used as a source of vapor, (6) high-resolution video camera, (7) water-cooled aluminum heat sink, (8) Peltier cooler TEC1-12715, (9) Peltier cooler TEC1-12706, and (10) 2 mm thick copper plate.
}
\label{electrolysis schematic}
\end{figure}

Electrolysis experiments were conducted in a custom-built cell powered by an autotransformer supplying a 50 Hz alternating current, which was rectified through a half-wave rectifier. Current and voltage were continuously monitored during the operation (Supplementary note 1.1). To ensure stable electrochemical conditions, an automated electrolyte refilling system maintained a constant liquid level throughout each run. See Fig \ref{electrolysis schematic}(a)  (Supplementary note 1.2). The cell employed a graphite anode (99.99\% C, 6 mm diameter) and a Ni cathode (99.5\% nominal Ni, 1 mm diameter) separated by 2 cm and immersed to a depth of 1 cm in 0.1M KHCO$_3$ aqueous electrolyte. The electrolyte was prepared using ultrapure Milli-Q water (resistivity: 18.4~\(\mathrm{M\Omega cm}\) at 300 K and ICP–MS analysis shows the absence of impurities down to the 1 ppb detection threshold) to minimize the risk of externally introduced contamination (Supplementary note 2). Electrolysis experiments were conducted at 5~V and 20~V, and eight samples were prepared at each of these voltages.

Following electrolysis, Ni cathodes were removed, rinsed with deionized water to eliminate surface KHCO$_3$, cut, and immediately transferred to the cloud chamber operating at approximately $-40^{\circ}\mathrm{C}$ \cite{andrade2024thermoelectric} (Fig\ref{electrolysis schematic}(b) and Supplementary note 3). The cloud chamber allowed direct observation of charged particle emissions from the reacted samples. The emissions from each cathode were recorded for 20 minutes using a high-resolution video camera, and the particle trajectories were analyzed frame by frame to extract track length and shape.

\section{Results and Discussion}
\begin{figure}[H]
    \centering
    \includegraphics[width=0.75\linewidth]{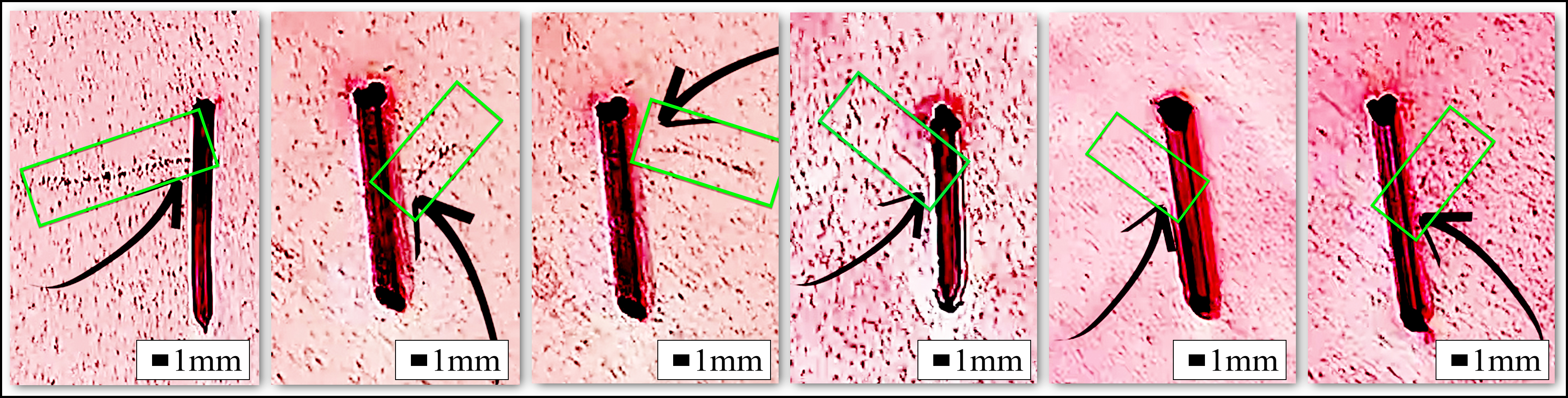}
    \caption{The track of each charged particle is contained within a green rectangle, with an arrow denoting its point of origin}
    \label{track collage}   
\end{figure}

Charged particle emissions were observed directly from the nickel cathodes post-electrolysis. See Fig.\ref{track collage}  (see also Supplementary Video~1). The detected tracks distinguished by their narrow width and continuous length (energy) distribution are consistent with $\beta$-like particle track characteristics\cite{leone2004note}. The condensation track lengths range from 0.6 mm to 16 mm. The mean detection rate for the 5~V samples is {\mathversion{normal}$0.6 \pm 0.1$} cpm and for the 20~V samples, {\mathversion{normal}$1.0 \pm 0.1$} cpm (Supplementary note 5), consistent across replicates and indicative of a reproducible, low-flux emission process. In contrast, the Ni controls prior to electrolysis did not yield any observable tracks (Supplementary Video 2), confirming that the emissions originated from the reacted Ni cathodes.

\begin{figure}[H]
    \centering
    \includegraphics[width=1\linewidth]{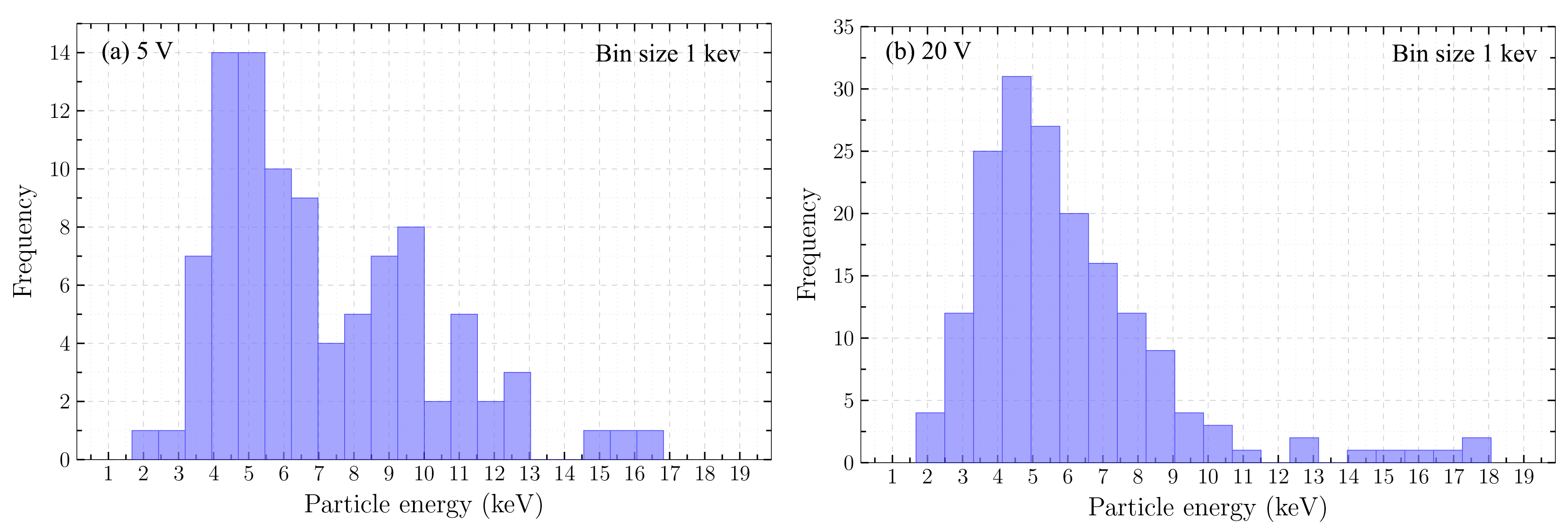}
    \caption{Energy distribution of observed charged-particle tracks (1 keV bin size) from samples prepared at (a) 5 V bias and (b) 20 V bias.}
    \label{energy histogram}
\end{figure}

The analysis of the track-lengths yielded the corresponding $\beta$-like particle energies of 2-18 keV. See Fig.\ref{energy histogram}. This is derived using the empirical range–energy relation\cite{cember2009introduction} (Supplementary note 5):

\begin{equation}
R = 0.407E^{1.38}
\label{eq:R}
\end{equation}
where R denotes the mean CSDA (continuous slowing-down approximation) range of electrons, expressed as mass thickness  (medium density multiplied by track-length in units of mg.cm$^{-2}$), representing the path-length over which a charged particle loses all of its kinetic energy $E$ (in keV ) in the medium. The medium density, corresponding to the supersaturated isopropyl-alcohol vapor in the air inside the cloud chamber, close to the samples,  was taken as $\rho = 1.1$~mg-cm$^{-3}$ (Supplementary note 4).

The observed spectrum, as per Fig.\ref{energy histogram}, is characterized by an endpoint energy of $\sim$18 keV and a peak energy of $\sim$5 keV, similar to the known beta decay spectrum of tritium. This alignment indicates that the detected $\beta$-like tracks can originate from the formation of tritium during electrolysis.\cite{notoya1994tritium,sankaranarayanan1996investigation}. The emission rate measured from our cloud chamber were {\mathversion{normal}$0.6 \pm 0.1$} cpm ($0.6 ~\text{Bq}$ from cathodes obtained after 5V electrolysis) and {\mathversion{normal}$1.0 \pm 0.1$} cpm ($1~\text{Bq}$  from cathodes obtained after 20V electrolysis).

A possible source of tritium could be contamination originating from the electrolyte. Natural light water contains only a trace amount of cosmogenic tritium, with typical concentrations of \(1.2\!-\!2.4\times10^{-12}\,\mathrm{ppm}\) in precipitation, corresponding to \(\sim0.14\!-\!0.29\,\mathrm{Bq\,kg^{-1}}\) \cite{WALLOVA2020106177,SUDPRASERT2025107604}. Assuming tritium loading from naturally tritiated water into the nickel lattice during 24\,h of electrolysis, the resulting tritium loading is bounded between $\sim1.2\times10^{-21}$ and $1.7\times10^{-18}$ tritium atoms per Ni atom, corresponding respectively to conservative (H/Ni $\approx10^{-3}$) and extreme upper-bound (H/Ni $\approx0.7$) hydrogen loading limits \cite{manchester2000}. These bounds yield absolute decay activities in the range $\sim10^{-9}$-$10^{-6}\,\mathrm{Bq}$ (Supplementary Note~7), which remain six to nine orders of magnitude below the experimentally observed emission rates.

To identify other possible origins of the observed low-energy $\beta$ tracks, we explored $\beta$-decaying isotopes with $Q \leq 100$~keV ( Supplementary note 7.3) and ruled them out as well. 

Contributions from $\alpha$-particles and $\gamma$-rays were also excluded. $\alpha$-particle induced tracks are characterized by substantially higher ionization density, greater track widths, and discrete energies, which are inconsistent with the observed tracks \cite{gyorfi2019dcc} (Supplementary note~6.1). The contribution of $\gamma$-ray interactions generating secondary electrons was excluded on the basis of an energy mismatch (Supplementary note~6.2).

Although diffusion-type cloud chambers are generally not regarded as ideal instruments for detecting low-energy $\beta$-particles owing to their inherently low detection efficiency (typically below 1\%) and strong dependence on thermal stability and vapor supersaturation, we achieved stable operation under optimized conditions. In our setup, we achieved stable conditions for approximately 60 minutes, during which the cold plate was maintained at approximately $-40^{\circ}\mathrm{C}$ and consistent isopropyl alcohol supersaturation was observed, as evidenced by the uniform droplet density within the chamber. Hence, with controlled operational conditions, it demonstrates reasonable sensitivity in the low-energy regime compared to conventional detection systems. Conversely, conventional scintillation detectors, including NaI(Tl)- activated crystals, liquid scintillation counters, and plastic scintillators, become ineffective at extremely low count rates due to their detection mechanisms and background noise, making reliable discrimination of true signals challenging (Supplementary note 8). In contrast, a cloud chamber provides direct, track-level visual evidence of individual charged-particle events.

Geometrical constraints, operational mechanisms, and user observation biases significantly limit the detection rate for the cloud chamber. These limitations result in an underestimation of activity and energy (Supplementary note 9).
 
Conventional empirical formulas for $\beta$-energy estimation were found to yield significant inaccuracies at such low energies, primarily because they neglect scattering effects. We have compared multiple energy estimations (Supplementary note 5). All of these empirical relations provided approximately the same decay energy profile. However, considering the measured track characteristics, the estimated energy range, and the sustained decay activity observed even after 6 months with only $\pm 1$ cpm variation from the initial measurements, the cumulative evidence supports that the detected $\beta$-tracks possibly originate from tritium decay.

\section{Conclusions}
Despite numerous suggestive indicators of nuclear activity in electrochemical systems, progress in this field has been hindered by the lack of direct experimental evidence of nuclear activity. We utilized custom-built cloud chambers, which have historically provided critical evidence of nuclear emissions, to investigate the emissions from Ni cathodes used in the electrolysis of light water. Post-electrolysis Ni cathodes emit low-energy $\beta$-like particles that are similar to the decay of tritium, suggesting tritium formation under ambient electrochemical conditions. Control experiments prior to electrolysis of Ni confirmed that particle emissions arose exclusively from electrochemical activation. The activity measured is orders of magnitude higher than possible through electrochemical loading of naturally occurring tritium in light water. These observations suggest that nuclear transformations can occur in condensed matter under electrochemical environments at energies far below conventional thresholds. This experiment highlights the cloud chamber as a sensitive method for probing extremely low activity and low-energy charged-particle emission, a regime in which conventional nuclear detection techniques exhibit very low efficiency and limited detection capability.

\section{Acknowledgments}
The authors thank Raviraj Singh Nehra and Jagdish Jangra for their valuable discussions and constructive feedback during manuscript preparation, which significantly improved the clarity and quality of the work.

\section{Author Contributions}
Shyam Sunder Lakesar designed and performed all experiments, carried out sample analyses, and drafted the manuscript. K. P. Rajeev and Raj Ganesh S. Pala critically reviewed and revised the manuscript. All authors discussed the results and approved the final version of the paper.

\end{document}